\documentclass[
aps,prl,
reprint,
showkeys,
twocolumn,
usenames,dvipsnames
]{revtex4-2}
\usepackage[utf8]{inputenc}
\usepackage[T1]{fontenc}
\usepackage{bm}

\usepackage{amsmath}
\usepackage{amsfonts}
\usepackage{slashed}

\usepackage{pgfplots}
\pgfplotsset{compat=1.14}
\usepackage{mathrsfs}
\usetikzlibrary{arrows}

\usepackage{booktabs}
\usepackage{multirow}

\usepackage{graphicx}
\usepackage{subfigure}
\usepackage{caption2}
\usepackage[colorlinks, citecolor=blue,anchorcolor=red,menucolor=red, linkcolor=red,filecolor=red,runcolor=red,urlcolor=blue,frenchlinks=red]{hyperref}

\usepackage{xcolor}

\usepackage{color}

\begin{document}

\title{Discontinuities of banana integrals in dispersion relation representation}
\author{Xu-Liang Chen$^1$}
\author{Peng-Fei Yang$^1$}
\author{Wei Chen$^{1,\, 2}$}
\email{chenwei29@mail.sysu.edu.cn}
\affiliation{$^1$School of Physics, Sun Yat-sen University, Guangzhou 510275, China \\ 
$^2$Southern Center for Nuclear-Science Theory (SCNT), Institute of Modern Physics, 
Chinese Academy of Sciences, Huizhou 516000, Guangdong Province, China}

\date{\today}

\begin{abstract}
We derive the discontinuities of banana integrals using the dispersion relation iteratively. We find a series of identities between the parameterized discontinuities of banana integrals (p-DOBIs). Similar to elliptic integrals, these identities enable the reduction of various p-DOBIs to be a linear combination of some fundamental ones. We present a practical application of p-DOBIs for deriving Picard-Fuchs operator. Then we establish the expression of generalized dispersion relation, which enables us to obtain the dispersion relation representation of arbitrary banana integrals. Moreover, we propose a hypothesis for generalized dispersion relation and p-DOBIs, which provides a simple way to calculate the discontinuities and transform dispersion relation representation to p-DOBIs. 
\end{abstract}

\maketitle

\textit{\textbf{Introduction.}} \textemdash \, Multi-loop Feynman integrals play a crucial role in the perturbative quantum field theory. Many of these integrals can be well understood by expressing in terms of Multiple Polylogarithms (MPLs)~\cite{Weinzierl:2022eaz,Duhr:2014woa,Bourjaily:2021lnz}. However, there are some physics processes involving complex elliptic Feynman integrals, which will go beyond the space of MPLs, for instance, the massive banana (or sunrise) integrals~\cite{bourjaily_functions_2022}. They have already given important contributions to the physics of Higgs decays, $\gamma\gamma$ production, $t\bar t$ production and so on. Many progresses have been made on computing such integrals analytically in recent years~\cite{adams_class_2018, adams_elliptic_2018, adams_iterated_2016, adams_two-loop_2014, adams_two-loop_2015, adams_varepsilon-form_2018, bloch_elliptic_2015, bloch_feynman_2015, bloch_local_2018, bonisch_analytic_2021, bonisch_feynman_2022, bourjaily_bounded_2019, broedel_analytic_2019, broedel_elliptic_2015, broedel_elliptic_2018, broedel_elliptic_2019, de_la_cruz_algorithm_2024, duhr_ice_2023, forum_symbol_2023, frellesvig_cuts_2021,  klemm_l-loop_2020, lairez_algorithms_2023, pogel_bananas_2023, pogel_feynman_2023, pogel_taming_2023, pogel_three-loop_2022, primo_maximal_2017,primo_maximal_2017-1, remiddi_differential_2016, remiddi_elliptic_2017, vanhove_differential_2021, vanhove_feynman_2018, vanhove_physics_2014}.

It is well-known that all elliptic integrals can be expressed linearly in terms of three fundamental integrals. Furthermore, the maximal cut of two-loop banana integrals is associated with elliptic integrals. In Ref.~\cite{remiddi_elliptic_2017}, a new class of functions called E-polylogarithm are defined as a generalization of MPLs and successfully applied to two-loop banana integrals. These functions exhibit similar structures to elliptic integrals. This inspires us to the intriguing question of whether similar properties hold for higher-loop banana integrals. 
In this letter, we provide a positive answer at the massive three-loop level. 

We define a novel class of functions as a natural parameterization of the discontinuities of banana integrals (p-DOBIs). Utilizing the technique of elliptic integrals, we obtain a series of identities between p-DOBIs. At three-loop level, we express all p-DOBIs as the linear combination of several fundamental ones. Furthermore, we demonstrate the application of p-DOBIs in deriving the Picard-Fuchs operators and reveal that the homogenous solutions of Picard-Fuchs equation can be given by slightly altering the definition of p-DOBIs.
	
To express an arbitrary banana integral with p-DOBIs, we establish the expression of generalized dispersion relation (GDR), which is crucial to deal with integrals involving high-power propagators. Furthermore, we propose a hypothesis to simplify the dispersion relation (DR) representation to p-DOBIs.

\textit{\textbf{The banana as iteration of the bubble.}} \textemdash \, 
An $L$-loop banana integral is defined as
\begin{equation}  
	\begin{gathered}
		B_{n_0n_1\cdots n_{L}}(q^2; \bm{m}  )
		= \int \prod_{i=1}^L  \frac{  d^D k_i }{ \text{i}   \pi^{\frac{D}{2}}  } 
		\frac{ 1   }
		{   \mathcal{D}_0^{n_0}   \mathcal{D}_i^{n_i}    }\, ,
	\end{gathered}
\end{equation}
where $\mathcal{D}_i = k_i^2 - m_i^2$ and $\mathcal{D}_0 =  \left( q-\sum_{i=1}^{L}k_i  \right)^2 -m_0^2$ are the inverse propagators, $  \bm{m} =\{m_0,m_1,\cdots,m_L\}$ is the set of masses, and $q$ is the external momentum. In this work, we shall use the following notations for convenience: $m_{ij}\equiv m_i+m_j$, $m_{0\cdots L}^2=(m_0+\cdots+m_L)^2$, $s_{ij} \equiv \sqrt{s_i}+m_j$ , $s_{i\bar{j}} \equiv \sqrt{s_i}-m_j$, $\lambda_n \equiv \lambda(s_{n},s_{n-1},m_{n}^2)$,  where $\lambda(x,y,z)= x^2 +y^2+z^2-2xy-2yz-2xz$, $s_0\equiv m_0^2$  , $\rho_{n_0n_1\cdots n_{L}}(q^2; \bm{m}  ) \equiv \pi^{-1} \text{Im}B_{n_0n_1\cdots n_{L}}(q^2; \bm{m}  )$. The discontinuity of the one-loop  $B_{11}(s;m_0,m_1)$ can be described by the spectral function 
\begin{equation} 
	\rho_{11}(s;m_0,m_1)\equiv\frac{1}{\pi}\text{Im}B_{11}(s;m_0,m_1) = 2 \lambda_1^{-\frac{1}{2}} .
\end{equation} We will limit our discussion to the $D=2-2\epsilon$ dimensions and $\epsilon^0$ order.

\par A two-loop banana integral can be represented as iteration of one-loop bubbles \cite{remiddi_differential_2016} 
\begin{equation}\label{b111}
	\begin{gathered}
		\begin{aligned}
			&B_{1^3}(q^2;  \bm{m}  )\equiv B_{111}(q^2;  \bm{m}  ) \\=& \int \frac{d^Dk_2 }{\text{i} \pi^{\frac{D}{2}}}   \int_{m_{01}^2}^{\infty} ds_1 \frac{  - \rho_{11}(s_1;m_0,m_1)  }{ (q-k_2)^2 - s_1  }   \frac{1}{ \mathcal{D}_2  }    \\
			=&  \int_{m_{01}^2}^{\infty} ds_1   \int_{s_{12}^2}^{\infty} ds_2    \frac{   \rho_{11}(s_1;m_0,m_1)  \rho_{11}(s_2;\sqrt{s_1},m_2)   }{ q^2-s_2} 
			\\
			=& 4  \int_{m_{012}^2}^{ \infty } ds_2      \int_{  m_{01}^2 }^{  s_{2\bar{2}}^2} ds_1  \frac{ \lambda_1^{-\frac{1}{2}}  \lambda_2^{-\frac{1}{2}}  }{q^2-s_2}, \,
		\end{aligned}
	\end{gathered}
\end{equation}
where the discontinuity of the propagator $(q^2-s_2 + i0^+)^{-1}$ is $ - \delta(s_2-q^2)$ for $q^2 \geq m_{012}^2$. Thus the two-loop DOBI is
\begin{equation} 
	\begin{gathered}
		\begin{aligned}
			\rho_{1^3} (s_2;  \bm{m} ) =  -4 \int_{  m_{01}^2 }^{  s_{2\bar{2}}^2} ds_1   \lambda_1^{-\frac{1}{2}}  \lambda_2^{-\frac{1}{2}}  .
		\end{aligned}
	\end{gathered}
\end{equation}
Similarly, the $L$-loop ($L \geq 2$) DOBI can be written as
\begin{equation}\label{Lloop}
	\begin{gathered}
		\begin{aligned}
			\rho_{1^L}(s_L; \bm{m} ) 
			=  (-1)^{L-1} 2^L  \prod_{i=2}^{L} \left(  \int_{m_{0\cdots (i-1)}^2}^{s_{i \overline{i}}^2}  \frac{ds_{i-1}}{ \sqrt{   \lambda_{i}  }}  \right)   \frac{1}{ \sqrt{   \lambda_{1}  }}  .
		\end{aligned}
	\end{gathered}
\end{equation}
Meanwhile, the real part is given by the Cauchy principal value of the integral
\begin{equation}\label{BLloop}
	\begin{gathered}
		\begin{aligned}
			\text{Re}B_{1^L}(q^2; \bm{m} ) 
			= \mathcal{P} \int_{m_{0\cdots i}^2}^{\infty}  ds_L \frac{ \rho_{1^L}(s_L; \bm{m} ) }{s_L-q^2} .
		\end{aligned}
	\end{gathered}
\end{equation}

One can prove that the results of Eq.~\eqref{BLloop} numerically agrees with the Bessel representation~\cite{groote_evaluation_2007,groote_numerical_2012,groote_coordinate_2019} for $q^2 < m_{0\cdots L}^2$ :
\begin{equation}
	\begin{gathered}
	B_{1^L}(q^2; \bm{m} ) =  \int_{0}^{\infty} dr \frac{r}{2} J_{0}(  r\sqrt{-q^2} ) \prod_{i=0}^{L} \left(  -2 K_0 (m_i r)   \right) ,  
	\end{gathered}
\end{equation}
where $J_0(x) $ is the first kind Bessel function and $K_0(x) $ is the second kind modified Bessel function.
 For $q^2>m_{0\cdots L}^2$, the results in  Eqs.~\eqref{Lloop}-\eqref{BLloop} numerically agree with the analytical continuation expression of Bessel representation ($q^2>0$):
\begin{equation}
	\begin{gathered}
B_{1^L}(q^2; \bm{m} ) = - \int_{0}^{\infty} dr \frac{r}{2} J_{0}(  r\sqrt{q^2} ) \prod_{i=0}^{L} \left(   i \pi H_0^{(2)}(m_i r)   \right)\, ,
	\end{gathered}
\end{equation}
where $H_0^{(2)}(x) $ is the second kind Hankel function, inspired by Ref.~\cite{zhang_note_2010}. For $0<q^2<m_{0\cdots L}^2$, the Hankel representation gives the same result with the Bessel representation.

\textit{\textbf{Identities between discontinuities of banana integrals.}} \textemdash \, Since the above results, we define the p-DOBIs as the following form
\begin{align}\label{empl_I2}
&\mathcal{I}_{k_1}^{a_1}(s_2 ; \bm{m}  ) =  \int^{s_{2\bar{2}}^2 }_{m_{01}^2} ds_1 (\lambda_1 \lambda_2)^{-\frac{1}{2}} (s_1-a_1)^{k_1}\, ,
\\ \nonumber
&\mathcal{I}_{k_1,\cdots,k_{n}}^{a_1,\cdots, a_{n}} (s_{n+1}; \bm{m}  )
\\
= & \int_{m_{0\cdots n}^2}^{s_{(n+1) \overline{(n+1)}}^2}  ds_{n}   \frac{  (s_{n}-a_{n})^{k_{n}}    }{ \sqrt{  \lambda_{n+1}  }  }   	\mathcal{I}_{k_1,\cdots,k_{n-1}}^{a_1,\cdots, a_{n-1}} (s_{n} ; \bm{m}  )\, , \label{empl_Ik}
\end{align}
where $a_1,\cdots,a_n$ are some useful parameters and we shall discuss them later. Comparing to Eq.~\eqref{Lloop}, $\mathcal{I}_{0,\cdots,0}^{a_1,\cdots, a_{L-1}} (s_{L}; \bm{m} )$ is proportional to $\rho_{1^L}(s_L; \bm{m}  )$ so that it can be used to describe DOBIs. In Ref.~\cite{remiddi_elliptic_2017}, $\mathcal{I}_{k_1}^{0}(s_2)$ has been adopted to calculate the two-loop equal mass banana integrals. Since $ \mathcal{I}_{k_1}^{a_1}(s_2) $ is an elliptic integral, all these integrals can be represented by three fundamental p-DOBIs. We will show that this can be generalized to the two-fold p-DOBIs $ \mathcal{I}_{k_1,k_2}^{a_1,a_2}(s_3) $ corresponding to the three-loop integrals.

We start from the following identity of $\mathcal{I}_{k_1}^{a_1}(s_2)$ \begin{align} \label{iddtk1}
\int^{s_{2\bar{2}}^2 }_{m_{01}^2} ds_1 \frac{d}{ds_1}  \left[  \sqrt{\lambda_1 \lambda_2} (s_1-a_1)^{k_1} \right] =0\, ,
\end{align}
where $a_1 \neq  s_{2\bar{2}}^2 ,m_{01}^2$. Noting that the upper and lower integral limits are the zero points of $\lambda_2$ and $\lambda_1$, respectively. 
Replacing $\lambda_{n}^{\frac{1}{2}} $ to $ \lambda_{n}^{-\frac{1}{2}} \lambda_{n} $ in Eq.~\eqref{iddtk1} and using Eq.~\eqref{empl_I2}, we find a series of identities between p-DOBIs and all these $\mathcal{I}_{k_1}^{a_1}(s_2)$ can be represented by three fundamental p-DOBIs 
\begin{equation}\label{wellknown-k1}
	\begin{gathered}
	 \{  \mathcal{I}_{-1}^{a_1}, \, \mathcal{I}_{0}^{a_1},\, \mathcal{I}_{1}^{a_1}    \}.
	\end{gathered}
\end{equation}

For $\mathcal{I}_{k_1,k_2}^{a_1,a_2}(s_3) $, we consider the similar equation as
\begin{align}\label{iddtk1k2}
\nonumber
& \int^{s_{3\bar{3}}^2}_{m_{012}^2} ds_2   \\ & \frac{d}{ds_2} \left[ 
\int^{ s_{2\bar{2}}^2 }_{m_{01}^2} ds_1    \frac{ \sqrt{\lambda_2\lambda_3 }}{  \sqrt{\lambda_1 } }   (s_1-a_1)^{k_1} (s_2-a_2)^{k_2} \right]
=0\, .
\end{align}
We can further find the identities for $\mathcal{I}_{k_1,k_2}^{a_1,a_2}(s_3) $ by using the identities of one-fold p-DOBIs obtained above. Setting $a_1=(m_0-m_1)^2=0, a_2 = (m_0-m_1+m_2)^2=1$ for the equal mass case $m_0=m_1=m_2=1$, the $ \mathcal{I}_{k_1,k_2}^{a_1,a_2}(s_3) $ with  $\{k_1 \geq -1,k_2\geq 0 \}$ can be represented by five fundamental p-DOBIs
\begin{equation}\label{wellknown-k1k2}
	\begin{gathered}
	 \vec{I}  =\{  \mathcal{I}_{-1,1}^{0,1}  ,   \mathcal{I}_{-1,2}^{0,1}    , \mathcal{I}_{0,0}^{0,1}      , \mathcal{I}_{1,0}^{0,1}      , \mathcal{I}_{1,1}^{0,1}    \}\, ,
	\end{gathered}
\end{equation}
in which the following well-known identity~\cite{remiddi_differential_2016, remiddi_elliptic_2017}
\begin{equation}\label{wellknown}
	\begin{gathered}
		 \mathcal{I}_1^0(s_2) = (1+\frac{1}{3}s_2)  \mathcal{I}_0^0(s_2)
	\end{gathered}
\end{equation}
is utilized to achieve Eq.~\eqref{wellknown-k1k2}. 
As an example, 
\begin{equation}
	\begin{gathered}
		\begin{aligned}
			\mathcal{I}_{-1,3}^{0,1} 
			=& \frac{1}{2}(4s_3 - s_3^2 )	\mathcal{I}_{-1,1}^{0,1}  
			+ \frac{1}{2}(s_3^2 + 8 s_3 + 32 )	\mathcal{I}_{0,0}^{0,1} 
			\\ - & \frac{1}{2}(11s_3+24)	\mathcal{I}_{1,0}^{0,1}  + 4 \mathcal{I}_{1,1}^{0,1}  + \frac{3}{2} s_3	\mathcal{I}_{-1,2}^{0,1}\, .
		\end{aligned}
	\end{gathered}
\end{equation}
One can easily check this result numerically.

To derive the differential equations for two-fold p-DOBIs, we study the following integral $\frac{d}{ds_3}\vec{I}$ by replacing $ \lambda_{3}^{\frac{1}{2}}$ to $ \lambda_{3}^{-\frac{1}{2}} \lambda_{3} $
\begin{equation}\label{eq4pf}
	\begin{gathered}
		\frac{d}{ds_3} \int^{s_{3\bar{3}}^2}_{m_{012}^2} ds_2   \int^{ s_{2\bar{2}}^2 }_{m_{01}^2} ds_1 
		\frac{ \sqrt{ \lambda_3}  }{  \sqrt{\lambda_1 \lambda_2 } }
		(s_1-a_1)^{k_1} (s_2-a_2)^{k_2}\, ,
	\end{gathered}
\end{equation}
which then can be represented with $\frac{d}{ds_3}  \mathcal{I}_{k_1,k_2}^{a_1,a_2}(s_3)$. On the other hand, Eq.~\eqref{eq4pf} can be expressed directly by $ \mathcal{I}_{k_1,k_2}^{a_1,a_2}(s_3)$ after finishing the derivatives. The differential equations $\frac{d}{ds_3}\vec{I}=\mathcal{A}^{(1)}  \vec{I}$ can be established with different $\{k_1,k_2\}$, in which $\mathcal{A}^{(1)} $ is the $5\times 5$ matrix with only simple poles 
\begin{widetext}
	\begin{equation}
		\begin{gathered}
			\mathcal{A}^{(1)} = \left( \begin{matrix}
				0&0&0&0&0 \\  \frac{1}{2} & 0 & - \frac{1}{2} &0 &0 \\ 0&0&0&0&0\\ 0&0&0&0&0\\  \frac{1}{2} & 0 & - \frac{3}{2} & \frac{1}{2} &0
			\end{matrix} \right) \frac{1}{s_3} 
			+
			\left( \begin{matrix}
				- \frac{1}{2} & \frac{1}{8} & \frac{1}{2} & - \frac{3}{8}  & 0  \\  0 & 0 & -4 &3 &-\frac{1}{2} \\ 0&-\frac{3}{16}  & -\frac{9}{4} &  \frac{27}{16}&  -\frac{3}{16}  \\ 0& -\frac{1}{4}&-3 & \frac{9}{4}& -\frac{1}{4}\\  0 & 0 & 0 & 0 & 0
			\end{matrix} \right) \frac{1}{s_3-1} 
			+
			\left( \begin{matrix}
				\frac{1}{2} & -\frac{1}{6} & - \frac{13}{6} & \frac{5}{3}  &  -\frac{1}{6}  \\   -\frac{1}{2} & \frac{1}{6} &  \frac{13}{6} & -\frac{5}{3} & \frac{1}{6} \\   -\frac{1}{2} & \frac{1}{6} &  \frac{13}{6} & -\frac{5}{3} & \frac{1}{6}  \\   -\frac{1}{2} & \frac{1}{6} &  \frac{13}{6} & -\frac{5}{3} & \frac{1}{6}  \\ \frac{1}{2} & -\frac{1}{6} & - \frac{13}{6} & \frac{5}{3}  &  -\frac{1}{6} 
			\end{matrix} \right) \frac{1}{s_3-4} 
			+
			\\
			\left( \begin{matrix}
				-\frac{1}{2} & \frac{1}{12} &  \frac{10}{3} & -\frac{7}{3}  &  \frac{5}{24}  \\  0 & \frac{2}{3} &  \frac{32}{3} & -\frac{26}{3} & \frac{2}{3} \\   0 & \frac{1}{24} &  \frac{2}{3} & -\frac{13}{24} & \frac{1}{24}  \\   0 & \frac{1}{6} &  \frac{8}{3} & -\frac{13}{6} & \frac{1}{6}  \\ 0 & \frac{4}{3} & \frac{64}{3} & -\frac{52}{3}  &  \frac{4}{3} 
			\end{matrix} \right) \frac{1}{s_3-16} 
			+
			\left( \begin{matrix}
				1 & -\frac{1}{24} &  -\frac{19}{6} & \frac{25}{24}  &  -\frac{1}{24}  \\  8 & -\frac{1}{3} & -\frac{76}{3} & \frac{25}{3} & -\frac{1}{3} \\   \frac{1}{2} & -\frac{1}{48} & -\frac{19}{12} & \frac{25}{48} & -\frac{1}{48}  \\   2 & -\frac{1}{12} &  -\frac{19}{3} & \frac{25}{12} & -\frac{1}{12}  \\ 16 & -\frac{2}{3} & -\frac{152}{3} & \frac{50}{3}  & -\frac{2}{3} 
			\end{matrix} \right)\, .
		\end{gathered}
	\end{equation}
\end{widetext}
The $n$-th differential equations $ \frac{d^n}{ds_3^n}  \vec{I}   = \mathcal{A}^{(n)}  \vec{I}$  can be easily derived from the first order ones.

These identities can be used to calculate the Picard-Fuchs operator for $ \mathcal{I}_{k_1,k_2}^{a_1,a_2}(s_3)$ 
\begin{equation}\label{PFO}
	\hat{L}(s_3) \mathcal{I}_{k_1,k_2}^{a_1,a_2}(s_3) \equiv \left(\sum_{i=0}^{n_{max}} P_i (s_3) \frac{d^i}{ds_3^i}\right)  \mathcal{I}_{k_1,k_2}^{a_1,a_2}(s_3) = 0,
\end{equation}
where $\hat{L}(s_3)$ is the Picard-Fuchs operator and $P_i (s_3)$ is a polynomial of $s_3$. The polynomial coefficients in $P_i (s_3)$ can be determined by letting the coefficient of each fundamental integral be zero for different values of $s_3$. Assuming $\hat{L}(s_3)$ be the third-order differential operator and the order of $P_i (s_3)$ be less than four, we obtain the well-known Picard-Fuchs operator for $\mathcal{I}_{0,0}^{a_1,a_2}(s_3) $~\cite{muller-stach_picard-fuchs_2014, vanhove_feynman_2018, lairez_algorithms_2023, broedel_analytic_2019,pogel_three-loop_2022,primo_maximal_2017,de_la_cruz_algorithm_2024}:
\begin{equation}
	\begin{gathered}
		\begin{aligned}
			\hat{L}(s_3) &= \left(s_3^4 - 20s_3^3 + 64s_3^2 \right)  \frac{d^3}{ds_3^3} 
			\\
			&+  \left( 6s_3^3 - 90 s_3^2 +192 s_3 \right)  \frac{d^2}{ds_3^2} 
			\\
			&+  \left( 7s_3^2 - 68 s_3 +64 \right)  \frac{d}{ds_3}  + (s_3 -4)\, .
		\end{aligned}
	\end{gathered}
\end{equation}
In principle, all $\mathcal{I}_{k_1,k_2}^{a_1,a_2}(s_3)$ can be expressed as series with the method of Frobenius.

We choose different integrating range of in Eq.~\eqref{empl_Ik} to define functions similar to p-DOBIs. If these functions still satisfy Eq.~\eqref{iddtk1k2} and Eq.~\eqref{eq4pf}, the identities and differential equations remain valid. Therefore, we can choose these functions as another two homogeneous solutions for Eq.~\eqref{PFO}, e.g. 
\begin{equation}
	\begin{gathered}
		\begin{aligned}
			\int^{  s_{33}^2 }_{ s_{3\bar{3}}^2 } ds_2   \int^{ s_{2\bar{2}}^2 }_{m_{01}^2} ds_1 
			\frac{ 1  }{  \sqrt{\lambda_1 \lambda_2 \lambda_3}    }\, ,
			\\
			\int^{  s_{33}^2 }_{ s_{3\bar{3}}^2 } ds_2   \int^{  s_{22}^2  }_{  s_{2\bar{2}}^2   } ds_1 
			\frac{ 1  }{  \sqrt{\lambda_1 \lambda_2 \lambda_3}    }\, .
		\end{aligned}
	\end{gathered}
\end{equation} 

We need eight fundamental p-DOBIs for the four-mass case, which are chosen as  $\vec{I}_{4m}  =\{\mathcal{I}_{-1,0}^{a_1,a_2}  , \mathcal{I}_{-1,1}^{a_1,a_2}  , \mathcal{I}_{-1,2}^{a_1,a_2}  , \mathcal{I}_{0,0}^{a_1,a_2}  , \mathcal{I}_{0,1}^{a_1,a_2}   , \mathcal{I}_{0,2}^{a_1,a_2}  , \mathcal{I}_{1,0}^{a_1,a_2}  , \mathcal{I}_{1,1}^{a_1,a_2} \}  $. The parameters are assigned as $a_1=(m_0-m_1)^2,\, a_2 = (m_0-m_1+m_2)^2$. The differential equations $ \vec{I}_{4m}$ and Picard-Fuchs operators can be derived following the same method as above. We obtain the same results for $\mathcal{I}_{0,0}^{a_1,a_2}$ as those in Refs.~\cite{de_la_cruz_algorithm_2024,lairez_algorithms_2023}. The parameters $a_1,a_2$ can also be other values, however, the pseudo-thresholds are recommended in our analyses.

\textit{\textbf{Generalized dispersion relation.}} \textemdash \, 
Integrals with high-power propagator are unavoidable issues to express arbitrary banana integral with p-DOBIs. However, it may lead to incorrect results by applying DR directly to integrals with high-power propagator. 

Considering the one-loop bubble with equal mass $m_0 = m_1 =1$ : 
\begin{equation}
B_{2,1}(q^2) = \int  \frac{d^2 k}{ \text{i}  \pi }  \frac{1}{\left[(q-k)^2-1 \right]^2(k^2-1)}\, ,
\end{equation}
the imaginary part is $\text{Im} B_{2,1}(s) = 2\pi s^{-\frac{1}{2}} (s-4)^{-\frac{3}{2}} $. Substituting $\text{Im} B_{2,1}(s)$ into DR
\begin{equation}\label{DR}
	\Pi(q^2) =\int_{s_N}^{\infty}  \frac{ds}{\pi}  \frac{  \text{Im}\Pi(s)  }{  s-q^2 }\, ,
\end{equation}
one will achieve the expression as
\begin{equation} 
	B_{2,1}(q^2) =    \int_{4}^{\infty} ds \frac{ 2  }{(s-q^2) s^{\frac{1}{2}}(s-4)  ^{\frac{3}{2}} }\, ,
\end{equation}
which is obviously divergent and thus incorrect. Such divergent problems were discussed very early in Ref.~\cite{Toll:1956cya}, and also in Ref.~\cite{Bargiela:2024hgt} recently by introducing subtraction terms. However, there is no explicit calculation provided in these works. In this work, we establish the generalized dispersion relation to deal with such divergent behavior at finite points and give their complete expressions. 

\begin{figure}[t!]
\centering
\includegraphics[width=6cm]{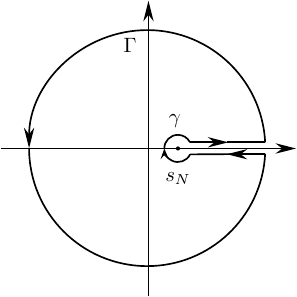}
\caption{Contour for integration of DR. } \label{FIG1}
\end{figure}

In general, a holomorphic function $\Pi(q^2)$ in the closed contour $C$ is expressed as
\begin{equation}  \label{eq2.1}
	\Pi (q^2) =   \left(\int_\Gamma +\int_\gamma \right) \frac{ds}{2\pi   \text{i} } \frac{ \Pi(s)}{s-q^2} +  \int_{s_N}^{\infty}  \frac{ds}{\pi}  \frac{ \text{Im} \Pi(s) }{s-q^2}\, ,
\end{equation}
in which there is a branch cut starting at $s_N$ on the real axis, as shown in Fig.~\ref{FIG1}.
The contribution of the small circle $\gamma$ is usually neglected in the expression of DR, since it appears only in integrals with high-power propagators. This contribution can be considered directly by the cumbersome calculations of the integral in the small circle~\cite{Palameta:2017ols}.

Similar to the subtractions on the large circle $\Gamma$, one can also do subtractions on the small circle by letting $\Pi_\gamma (q^2) = (q^2-s_N)^{n_\gamma} \Pi(q^2)$ to avoid the divergences at $s_N$. Substituting $\Pi_\gamma (q^2)$  into DR with subtractions, we find the GDR at the subtraction point $a$:
\begin{equation} \label{GDR1}
	\begin{gathered}
		\begin{aligned}
			\Pi(q^2) = \sum_{i=0}^{n_\Gamma-1} 
			\frac{ (q^2-a)^{i} \left[ (q^2-s_N)^{n_\gamma}  \Pi(q^2)   \right]   ^{(i)}  |_{q^2=a}  }{  \Gamma(i+1) (q^2-s_N)^{n_\gamma} }  \\+
			\int_{s_N}^{\infty} \frac{ds}{\pi} \frac{ \text{Im} \Pi(s) }{s-q^2} 
			\left(  \frac{q^2-a}{s-a}  \right)^{n_\Gamma} \left(  \frac{s-s_N}{q^2-s_N}   \right)^{n_\gamma},
		\end{aligned}
	\end{gathered}
\end{equation}
where $n_\Gamma, n_\gamma$ are the subtraction times on the large circle and small circle, respectively. Since $n_\Gamma$ is determined by $\Pi_\gamma (q^2)$ rather than $\Pi(q^2)$, one should firstly find $n_\gamma$ in GDR similar with the way to find $n_\Gamma$ in DR. 
The expression of GDR in Eq.~\eqref{GDR1} can be naturally degenerated to DR for $ n_\gamma=0$. 


For the equal mass one-loop bubble, the complete expression of $B_{2,1}(q^2)$ can be obtained from GDR in Eq.~\eqref{GDR1} for $n_\gamma=1, n_\Gamma=1$ and subtraction point at $a=0$
\begin{equation}\label{b21}
	B_{2,1} (q^2) =   \frac{ 2}{ q^2-4  }  + \int_{4}^{\infty} \frac{ds}{\pi} \frac{\text{Im} B_{2,1}(s) }{s-q^2}   \frac{q^2}{s}   \frac{s-4}{q^2-4} .
\end{equation}
This expression numerically agrees with the result in Mellin-Barnes representation~\cite{Boos:1990rg}. 

The result of GDR is very useful to deal with the divergent banana integrals with high-power propagators. Using the same method in Eq.~\eqref{b21}, we can write the discontinuity of the two-loop $B_{211}(s_2;1,1,1)$ as
\begin{align} \label{rho211}
\nonumber &	\rho_{211}(s_2;1,1,1)
\\ \nonumber
=& 2 \rho_{11}(s_2;2,1) + \int_{4}^{\infty} ds_1 \frac{4}{s_1}  \rho_{21}(s_1;1,1)   \rho_{11}(s_2;2,1)  
\\
-&  \int_{4}^{s_{2\bar{2}}^2} ds_1 \rho_{21}(s_1;1,1)   \rho_{11}(s_2;\sqrt{s_1},1)\, ,
\end{align}
in which the GDR of $B_{2,1} (q^2)$ in Eq.~\eqref{b21} is adopted. 
In Eq.~\eqref{rho211}, the third term, which is divergent, comes from only the DR representation. The first two terms come from the GDR, in which the second term is also divergent. However, the divergences in the second and third terms cancel out exactly to each other so that $\rho_{211}(s_2;1,1,1)$ is finite. The GDR representation of $B_{211}(q^2;1,1,1)$ is obtained by substituting Eq.~\eqref{rho211} back into Eq.~\eqref{GDR1} with $n_\Gamma = n_\gamma = 1$ and $a=0$:
\begin{align} \label{s211}
\nonumber
B_{211}(q^2;1,1,1&)
=-\frac{9 B_{211}(0;1,1,1) }{q^2-9} 
\\ & +  \int_{9}^{\infty} ds_2 \frac{\rho_{211}(s_2;1,1,1)}{s_2-q^2} \frac{q^2}{s_2} \frac{s_2-9}{q^2-9} ,
\end{align}
which is convenient for numerical integration. However, one should note that the real part contains the Cauchy principal value integral for $q^2>9$, which may  reduce numerical accuracy.

Using the result of GDR, an arbitrary banana integral can be represented as an iteration of bubbles. Meanwhile, it is convenient to find the discontinuities.

\textit{\textbf{ Hypothesis about GDR and p-DOBIs. }} \textemdash \, In this section, we try to discuss the integrals arising in GDR. Let's discuss the divergence of the third term from DR in Eq.~\eqref{rho211} 
\begin{equation}
	\begin{gathered}\label{rho211last}
		\begin{aligned}
		 & -\int_{4}^{s_{2\bar{2}}^2} ds_1 \rho_{21}(s_1;1,1)   \rho_{11}(s_2;\sqrt{s_1},1)  
		 \\
		 =& -4 \int_{4}^{s_{2\bar{2}}^2} ds_1 (\lambda_1 \lambda_2)^{-\frac{1}{2}} (s_1-4)^{-1}
= -4 \mathcal{I}_{-1}^4(s_2)\, .
		\end{aligned}
	\end{gathered}
\end{equation}

We use the same trick of Eq.~\eqref{iddtk1} to deal with $ \mathcal{I}_{-1}^4(s_2)$. The equation should be modified as the following for the parameter $a_1=m_{01}^2=4$ at the normal threshold:
\begin{equation}\label{iddtk2}
	\begin{gathered}
		\begin{aligned}
			 &\left[  \sqrt{\lambda_1 \lambda_2} (s_1-a_1)^{k_1} \right] \bigg|_{s_1=m_{01}^2}^{s_1 = s_{2\bar{2}}^2}
			 \\= &  \int^{s_{2\bar{2}}^2 }_{m_{01}^2} ds_1 \frac{d}{ds_1}  \left[  \sqrt{\lambda_1 \lambda_2} (s_1-a_1)^{k_1} \right].
		\end{aligned}
	\end{gathered}
\end{equation}
For $k_{1}=-1$,
\begin{align}\label{iddtk1u}
\nonumber \frac{\sqrt{\lambda_1 \lambda_2} }{(s_1-4)}  \bigg|_{s_1=4}^{s_1 = s_{2\bar{2}}^2}
=& \mathcal{I}_2^4(s_2) -2(s_2-1)(s_2-9) \mathcal{I}_{-1}^4(s_2)
\\ &
 - (s_2 - 5)\mathcal{I}_1^4(s_2)\, .
\end{align}
The boundary term at the left hand side (LHS) shows how $\mathcal{I}_{-1}^4(s_2)$ diverges. In general, one can solve Eq.~\eqref{iddtk1u} and substitute $-4 \mathcal{I}_{-1}^4(s_2)$ back into Eq.~\eqref{rho211}. 

On the other hand, we find an alternative way to determine the DOBI ($s_2>9$):
\begin{align}\label{211hyp}
\nonumber
			&	\rho_{211}(s_2;1,1,1)
			\\ \nonumber
			= &-4\left[ \mathcal{I}_{-1}^4(s_2) + \frac{1}{2(s_2-1)(s_2-9)} \frac{\sqrt{\lambda_1 \lambda_2} }{(s_1-4)}  \bigg|_{s_1=4}^{s_1 = s_{2\bar{2}}^2}  \right] 
				\\
			= & \frac{2(s_2 - 5)\mathcal{I}_1^4(s_2) -2 \mathcal{I}_2^4(s_2) }{(s_2-1)(s_2-9)}\, ,
		\end{align}
in which we have used Eq.~\eqref{iddtk1u} to obtain the finite result in the last step. 

It shows that we can simply calculate the term from DR iterations and reduce it to p-DOBIs. Using the identities between p-DOBIs, we obtain the final results containing only finite p-DOBIs. The boundary term with divergence can be directly removed. This is our hypothesis about GDR and p-DOBIs. If this hypothesis is true, one can in principle express any banana integral with p-DOBIs.

Using this hypothesis, we have calculated the DOBIs $\rho_{211}(s_2;1,2,3)$, $\rho_{311}(s_2;1,1,1)$ , $\rho_{2111}(s_3;1,1,1,1)$, and our calculations are in good agreement with the results of GDR.  

According to Eq.~\eqref{wellknown}, only two independent one-fold p-DOBIs are needed in the equal mass case. It is well-known that $B_{1^3}(q^2;1,1,1)$ and $B_{211}(q^2;1,1,1)$ are two master integrals for equal mass sunrise integral at top sector~\cite{Tarasov:1997kx,Laporta:2004rb}, which is actually suggested in Eq.~\eqref{211hyp}. It is possible to extend this idea to high-loop or multi-mass cases.

\textit{\textbf{Summary.}} \textemdash \, In this letter, we define the p-DOBIs by parameterizing the DOBIs obtained in DR representation. The identities among p-DOBIs enable a variety of p-DOBIs to be expressed linearly in terms of some fundamental p-DOBIs. The p-DOBIs identities are useful tools for elliptic Feynman integrals. Similar to the IBP identities revealing the structure of Feynman integrals, the p-DOBIs identities can unveil the underlying structure of the maximal cut. We have demonstrated several potential applications of this approach, which is capable of handling multi-scale integrals and has the potential for extension to higher loops. The properties of p-DOBIs make it possible to generalize the definition of E-polylogarithm functions to three loops, which is a powerful tool for exploring higher-order $\epsilon$ expressions of banana integrals ~\cite{remiddi_elliptic_2017}.

The DR iteration will give incorrect divergent result for integrals involving high-power propagators. We establish the GDR to solve such problems by considering the subtractions on both small and large circles of the integration contour. 
The GDR iterations express DOBIs as integrals over tree-level amplitude. This recursive approach may have more applications, such as the integrated unitarity equations applied to the 4-loop 4-point massless planar ladder Feynman integrals~\cite{Bargiela:2024hgt}. 
We further find that one only needs to consider the term from DR iterations. Assuming the hypothesis about GDR and p-DOBIs, we can determine the multi-loop DOBIs in terms of finite p-DOBIs by ignoring the divergent boundary term.
%

%
%

\textit{\textbf{Acknowledgments.}} \textemdash \, Xu-Liang Chen thanks Zhi-Zhong Chen for useful discussions. This work is supported by the National Natural Science Foundation of China under Grant No. 12175318, the Natural Science Foundation of Guangdong Province of China under Grant No. 2022A1515011922.

 \bibliographystyle{apsrev4-2}

\end{document}